# NEURO HAND: A weakly supervised Hierarchical Attention Network for interpretable neuroimaging abnormality Detection

David A. Wood[1]

[1] School of Biomedical Engineering and Imaging Sciences, King's College London, Rayne Institute, 4th Floor, Lambeth Wing, London SE17 7EH, United Kingdom

## Abstract

Clinical neuroimaging data is naturally hierarchical. Different magnetic resonance imaging (MRI) sequences within a series, different slices covering the head, and different regions within each slice all confer different information. In this work we present a hierarchical attention network for abnormality detection using MRI scans obtained in a clinical hospital setting. The proposed network is suitable for non-volumetric data (i.e., stacks of high-resolution MRI slices), and can be trained from binary examination-level labels. We show that this hierarchical approach leads to improved classification, while providing interpretability through either coarse inter- and intra-slice abnormality localisation, or giving importance scores for different slices and sequences, making our model suitable for use as an automated triaging system in radiology departments.

1. ## Introduction

Deep learning-based computer vision systems hold promise for automatically triaging patients in hospital radiology departments. In the UK, for example, with a 4.6% increase in brain magnetic resonance imaging (MRI) scans performed in the last 12 months alone (NHS, 2019), and with an increase in the time taken to report out-patient MRI scans every year since 2012, an automated triage mechanism to identify abnormalities at the time of imaging, and thereby allow prioritised scan reporting, is urgently needed. Such a mechanism would potentially allow early intervention to improve short- and long-term clinical outcomes. Assuming that a first-generation system will operate by assisting real-time radiologist review, any prospective model must provide a quickly visualizable justification for its decision. Interpretability would also be essential to engender radiologist confidence and support clinical trials of second-generation autonomous systems (Booth et al., 2020). Ideally, this visualization would take the form of abnormal tissue segmentation, with the model outputting pixel-level probabilities in addition to accurate scan classification (i.e., normal vs. abnormal). However, training such a model by supervised learning requires large numbers of manually segmented images which are often not readily available. One approach to circumvent this bottleneck is to directly apply in clinical settings those models trained on curated open-access data collections that do have segmentation labels, such as the Brain Tumour Segmentation Challenge (BRATS) (Menze et al., 2015), or Ischemic Stroke Lesion Segmentation Challenge (ISLES) (Winzeck et al., 2018) datasets. However, these off-the-shelf models, being trained on standardized and often heavily pre-processed (i.e., skull stripped, spatially co-registered, isotropic) volumetric images, often suffer from domain shift; in other words, they fail to generalise to less homogeneous datasets such as the wide range of MRI scans generated at hospitals.

An alternative approach is to develop a model trained on these less homogeneous hospital datasets using simple classification information (i.e., normal vs. abnormal scan) in order to coarsely localise, rather than segment an abnormality (Wood et al., 2020), (Wood et al., 2021). Localisation of this kind, although not suitable for precision applications such as computer guided surgery and planning, is ideal for triage systems where the priority is to quickly identify and present the location of an abnormality for radiologist review (Din et al., 2023), (Agarwal et al., 2023), (Wood et al., 2022).

In this work we present a hierarchical attention model for automated abnormality detection from weak supervision labels. We characterise weak as being at the series-level. An MRI series is the entire set of MRI scans, incorporating multiple sequences (such as $T_1$-weighted, $T_2$-weighted, diffusion-weighted sequences), obtained during a patient's scanning session. Built around nested long short-term memory (LSTM) units and convolutional neural networks (CNNs), the proposed network is suitable for non-volumetric data (i.e., stacks of high-resolution MRI slices), and can be trained on minimally processed images extracted from hospital picture and archiving systems (PACS) and labelled using a recently developed radiological report language model (ALARM) (Wood et al., 2020). We show that this hierarchical approach leads to improved classification, while coarsely localising the inter- and intra-slice abnormality. The proposed approach is general, and would allow integration of other information relating to a study e.g., data from different imaging modalities (e.g., computed tomography (CT) or positron emission tomography (PET)) or even non-imaging data such as patient clinical history, all of which are highly desirable in a clinical setting (Booth et al., 2020). We have demonstrated this integration by incorporating multiple MRI sequences and have shown that such a strategy outperforms sum-pooling models and recurrent networks without attention, while providing importance scores for each sequence and slice.

2. Related work

Weakly supervised abnormality detection has attracted considerable interest in recent years. To date, most approaches have been based around class activation mapping (CAM) (Zhou et al., 2015), whereby candidate regions of interest generated using fully convolutional networks are processed to generate pixel-level segmentation maps (Feng et al., 2017), (Wei et al., 2017), (Izadyyazdanabadi et al., 2018), (Wu et al., 2019). One limitation of this approach is the requirement of slice- rather than series-level labels, meaning that all slices from each sequence used in a training set need to be manually labelled for the presence or absence of an abnormality or lesion. This makes the construction of large, labelled datasets considerably more time-consuming and expensive. A further shortcoming is the implicit treatment of slices as being independent of each other, thereby failing to leverage inter-slice spatial dependencies for abnormality detection.

Our work builds on that of (Poudel et al., 2016) and (Cai et al., 2018), treating variable-length stacks of MRI slices as correlated information and processing these data using re current convolutional networks. Crucially, we relax the requirement for pixel-level labels by incorporating a hierarchical attention mechanism - first introduced for language modelling (Yang et al., 2016) - to exploit the natural hierarchies present in neuroimaging data. In this way, our model is similar to (Zhang et al., 2017), (Yan et al., 2019), (Cole et al., 2020), and (Wood et al., 2019) who used visual attention to provide a form of model interpretability for medical image analysis. To our knowledge, however, this is the first demonstration of using hierarchical attention for weakly-supervised neurological abnormality detection.

### 3. Methods
### 3.1 Data

The UK National Health Research Authority and Research Ethics Committee approved this study. All 126,556 adult (≥ 18 years old) MRI head scans performed at KCH hospital between 2008 and 2019, were used in this study. MRI scans were obtained on Signa 1.5 T HDX General Electric Healthcare; or AERA 1.5T, Siemens, Erlangen, Germany. Using the ALARM radiology report classifier described in (Wood et al., 2020) and (Wood et al., 2022), all examinations were assigned a binary label, corresponding to the presence or absence of an abnormality predicted on the basis of the accompanying free text neuroradiology report describing the study. The classification accuracy of this model is 99.4%, so this labelling procedure is considered reliable. A subset of 600 abnormal examinations were then selected for inclusion using an open-source annotation tool (available at https://github.com/tomvars/sifter, see (Wood et al., 2020)). Because the hospital dataset consisted of MRIs obtained at different stages of the patient pathway (including initial diagnostic imaging, pre-surgical planning, immediate post-surgical assessment, and chemoradiotherapy response assessment over a longer period of follow-up), the abnormalities incorporated were heterogeneous, including tumours at diagnosis, resection cavities after surgery, and post-treatment related effects at follow-up (Fig. 1). As such, this dataset provided our model with abnormalities to train on that varied in size and MRI signal characteristic. 600 series were then randomly selected from a large subset labelled as normal to make a combined balanced dataset of 1200 examinations.

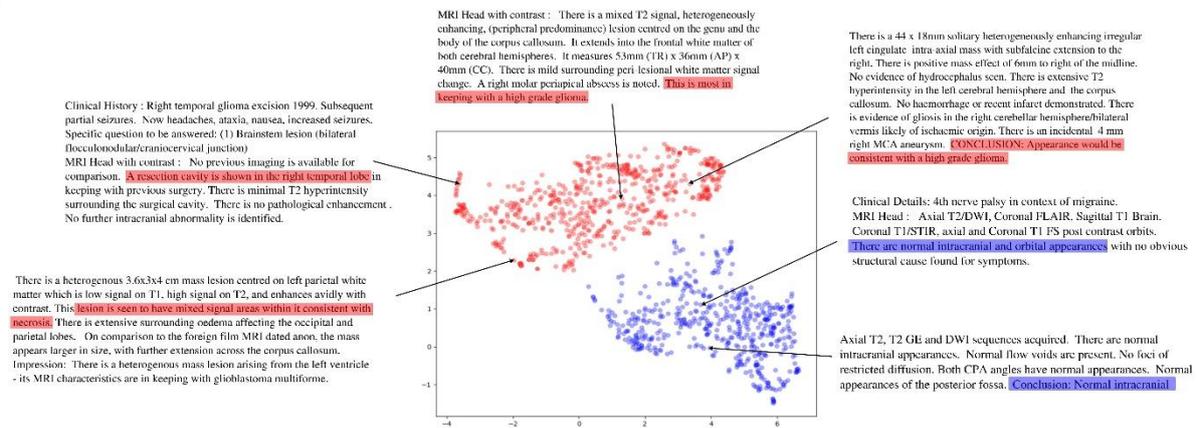

**Figure 1:** *Visualization of the contextualized report embeddings for normal (blue) and abnormal (red) cases captured using an open source 'lasso' tool (Wood et al., 2020b) for use in this study.*

### 4. Hierarchical attention network

In this work we present two hierarchical attention networks (HAN): one taking as input intra-slice regions spanning multiple image slices, and another taking as input individual slices spanning multiple imaging sequences.

### 4.1 Intra- and inter-slice HAN

Our intra- and inter-slice HAN (hereafter patch-HAN) was designed for abnormality detection using a collection of slices from a single MRI sequence (Fig. 2). The model treated each scan as being hierarchically composed of a number of slices, $N_{slice}$, with each slice itself composed of a number of local regions (patches). The number of patches, $N_{patch}$, was a model hyper-parameter, with lots of small patches improving abnormality localisation, but increasing the computational cost. A CNN looped through all patches in a given slice, processing each sequentially before passing its output to a bidirectional LSTM unit which built up an internal representation of what it had seen in that slice. The LSTM had $N_{patch}$ outputs per slice, and these were passed to an attention network to calculate the importance of each patch. The resulting weighted sum of patches become the representation for that slice. Following (Yang et al., 2016) the attention weights were computed as follows:

$$u_{it} = \tanh(W_w h_{it} + b_w)$$

$$\alpha_{it} = \frac{exp(u_{it}^T u_w)}{\sum_t exp(u_{it} u_w)}$$

$$s_i = \sum_t \alpha_{it} h_{it} \qquad \text{(Eqn. 1)}$$

where $s_i$ is the representation for the *i*'th slice, and α$_{it}$ is the weighting of the *t*'th patch representation in slice *i*, h$_{it}$. The new parameters to learn were therefore a context vector $u_w$, a matrix $W_w$ and a bias $b_w$ for each attention module. This procedure was repeated for all slices, with a second bidirectional LSTM taking these slice representations, outputting N$_{slice}$ hidden states (the number of slices) which were sent to another attention module to compute the importance of each slice. Finally, a weighted combination of slices (i.e., a weighted combination of a weighted combination of local patches) was sent to a single layer classifier which was trained by minimizing the binary cross-entropy at the level of series labels (i.e., all slices in a series have the same label; 1 for abnormal, and 0 for normal)

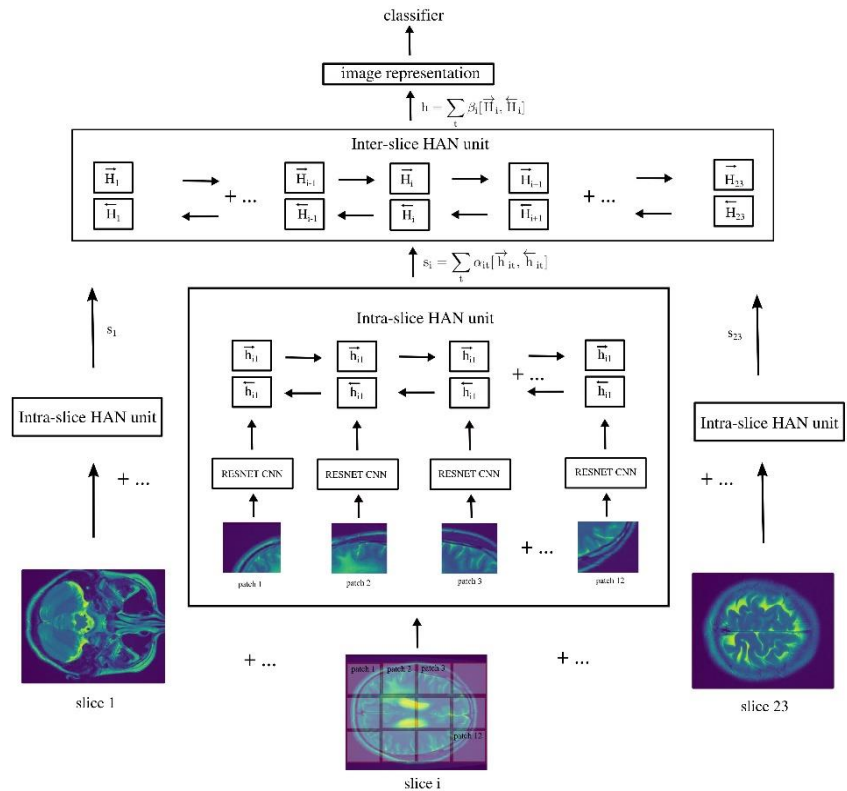

**Figure 2:** *Patch-HAN. The network builds up an image representation as a weighted combination of slice representations, with the sequence representations built from weighted combinations of patch representations.*

### 4.2 Inter-slice and sequence HAN

Our inter-slice and sequence HAN (hereafter sequence-HAN) was designed for abnormality detection using a collection of image slices from multiple MRI sequences. Like the intra- and inter-slice network, it built up a hierarchical data representation for abnormality classification, but this time at the series level. This was achieved by first building up individual sequence representations using separate LSTM units, CNNs, and slice-level attention modules. A sequence-level attention module then weighed the importance of each sequence representation, and the resulting series-level representation was again sent to a single layer classifier (Fig. 3).

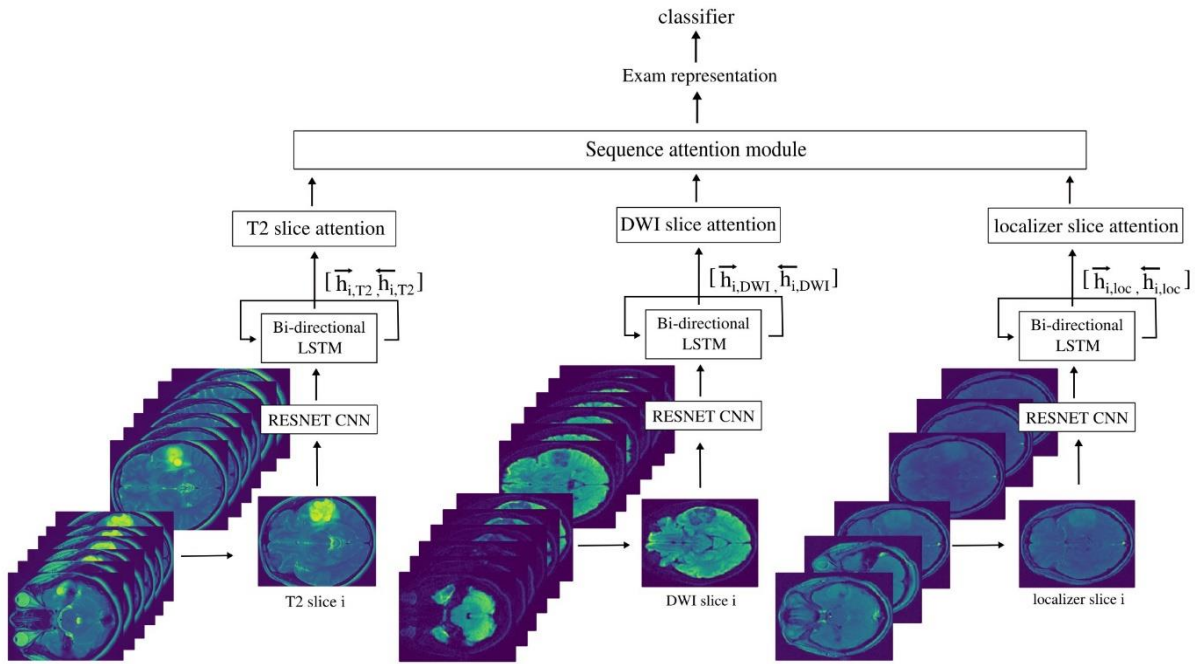

**Figure 3:** *Sequence-HAN. The network builds up a series representation as a weighted combination of sequence representations, where each sequence representation is constructed from a weighted combination of slice representations.*

### 5. Experiments

In this work we presented two proof-of-principle experiments to demonstrate the general utility of treating clinical neuroimaging data hierarchically - one using sequence-HAN to determine the most informative sequence(s) for this dataset, and a second using patch-HAN to perform more granular abnormality detection using this sequence. For the first experiment we included axial $T_2$-weighted images of size (512 x 512) x 23 slices, axial diffusion-weighted images (DWI) of size (256 x 256) x 7 slices, and axial localizer images of size (256 x 256) x 7 slices, as these images were common to nearly all clinical examinations. We split this dataset into training/validation/test sets of sizes 800, 200, 200, respectively, where each instance contains a stack of slices for each of the three sequences. Because this is real-world hospital data, multiple series for patients are common and images from these separate visits are likely to be highly correlated. To avoid this form of data leakage we preformed the split at the level of patients so that no patient that appeared in the training set appeared in the validation or test set. The images were then minimally pre-processed, with each pixel normalized to the slice mean, with unit variance. No skull-stripping or co-registration was performed reducing the complexity of the process and computational burden. For all experiments, the attention context vectors, matrices, and biases were initialized from a zero-centred normal distribution with variance σ = 0.05. In all experiments the CNNs were 18-layer ResNet networks, warm started using values pre-trained on ImageNet (Deng et al., 2009). The networks were trained for 15 epochs (with early stopping) using ADAM (Kingma and Ba, 2014) with initial learning rate 1e-4, decayed by 0.97 after each epoch, on a single NVIDIA GTX 2080ti 11 GB GPU. Minimal hyperparameter tuning (in this case learning rate and LSTM dimensionality) was performed on the validation set, and the model with the best classification accuracy was used to determine the final model performance on the balanced independent test set, For the results that we present, all LSTM hidden units had a dimension of 512, and the patch size for patch-HAN was 150 x 150. To benchmark patch-HAN, we train a baseline network which puts equal weighting on each slice and patch, and a second which processed each slice and patch independently and performed sum pooling (i.e., a fully convolutional model with no recurrent network). For the inter-sequence HAN we trained a convolutional model which performed sum-pooling over slices and sequences, as well as a recurrent-based network which put equal weighting on each sequence and slice.

### 5.1 Results

The classification performance of sequence-HAN, along with that of the two multi-sequence baseline architectures, appears in Table 1. Our model outperforms all simpler multi-sequence networks, achieving a classification accuracy of 95.5%, illustrating the value of treating different sequences and slices hierarchically. By analysing the attention weights of our model, the most informative sequence and/or slice can be determined for a particular study. Figure 4 shows the sequence weights for two examples from the test set, as well as the weights of each slice for the most informative sequence (in both cases $T_2$-weighted). These examples were representative; in general, the model put most weight on the $T_2$-weighted sequence for classification, with average scores of 0.81, 0.13, and 0.06 across the whole test set for $T_2$-weighted, DWI, and localizer sequences respectively. Good agreement between the slice attention weights and the spatial distribution of the abnormalities was demonstrated.

| Model | Accuracy (%) | Sensitivity (%) | Specificity (%) |
|---|---|---|---|
| **Multi-sequence** | | | |
| Sequence-HAN | **95.5** | **94** | **97** |
| No attention | 92.5 | 91 | 94 |
| Sum-pool slices and sequences | 89.5 | 88 | 91 |
| **Single sequence** | | | |
| Patch-HAN | **98.5** | **98** | **99** |
| No attention | 91.5 | 90 | 93 |
| Sum-pool slices and sequences | 87.5 | 85 | 90 |

**Table 1:** *Performance of patch- and sequence-HAN on the abnormality classification task, along with that of the baseline architectures. Best performance in bold.*

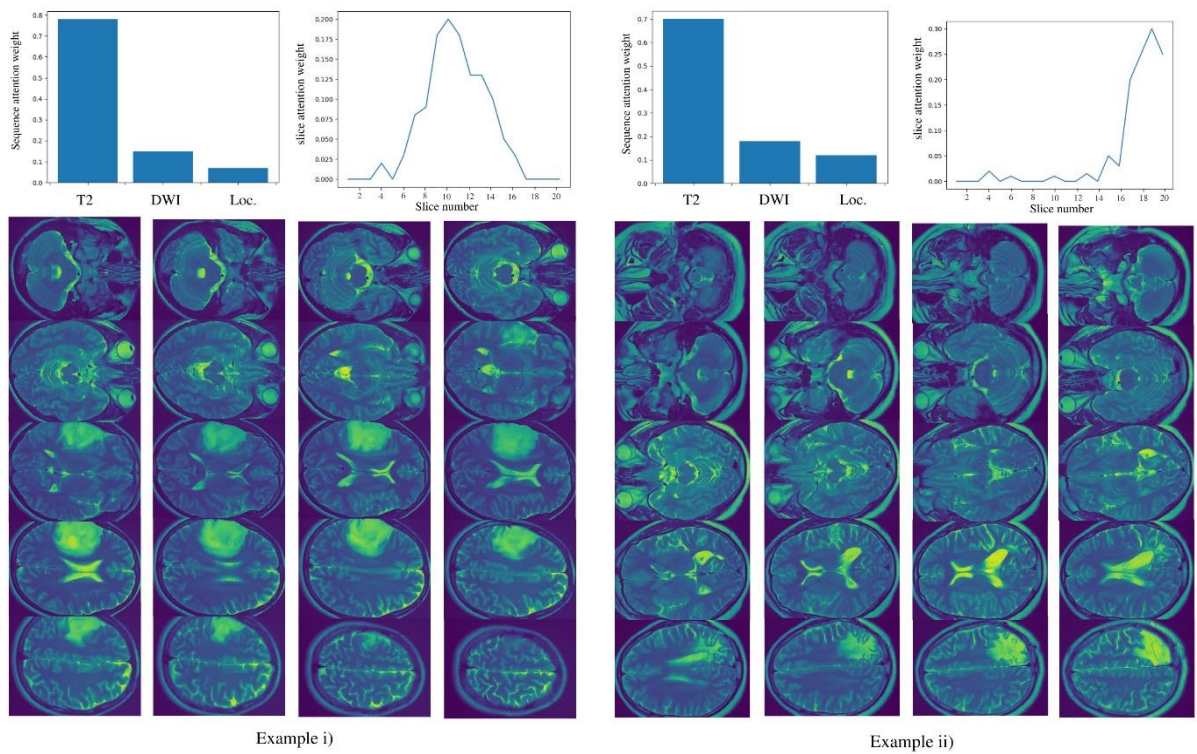

**Figure 4:** *Sequence and slice attention scores for two test set studies. Also included are the raw images for the most informative sequence ($T_2$). In both cases the slice attention weights closely match the spatial distribution of the abnormality.*

Figure 5 presents the most informative slices for each modality for two additional test set examples - the sequence attention weights closely matched the visibility of the abnormality on the corresponding sequence.

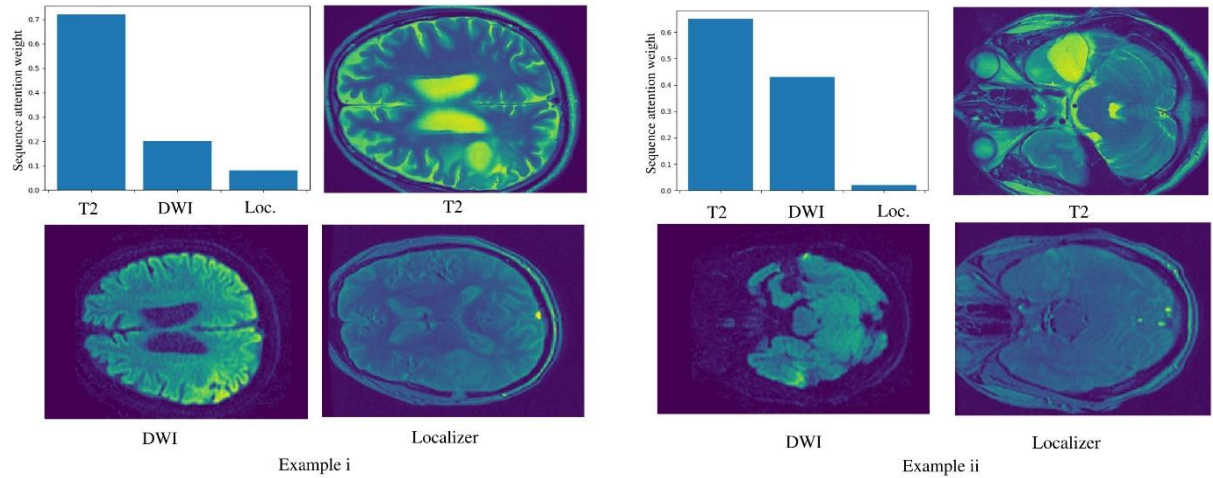

**Figure 5:** *Sequence scores for two test set studies. Also included are the most informative slices for each sequence. In both cases the sequence attention weights closely match the visibility of the abnormality on the corresponding sequence.*

Given the discriminative power of T2-weighted images for this task, we trained the patch-HAN using only these images to additionally localise abnormalities within MRI slices; the results also appear in Table 1. The hierarchical model outperforms all single-sequence baseline architectures. Figure 6 displays the slice and patch attention weights for two test set examples. In both cases, the top left plot displays the distribution of attention weights over each slice and the top right plot shows the most informative patch for the most informative slice. For reference the raw slices are shown as well. Again, the slice attention scores broadly agree with the spatial distribution of the abnormality across slices. The classifier seems able to classify and localise abnormalities across the entire brain volume (Fig 7), and even gave outputs with multi-modal distributions in cases where multiple, spatially separate abnormalities were present.

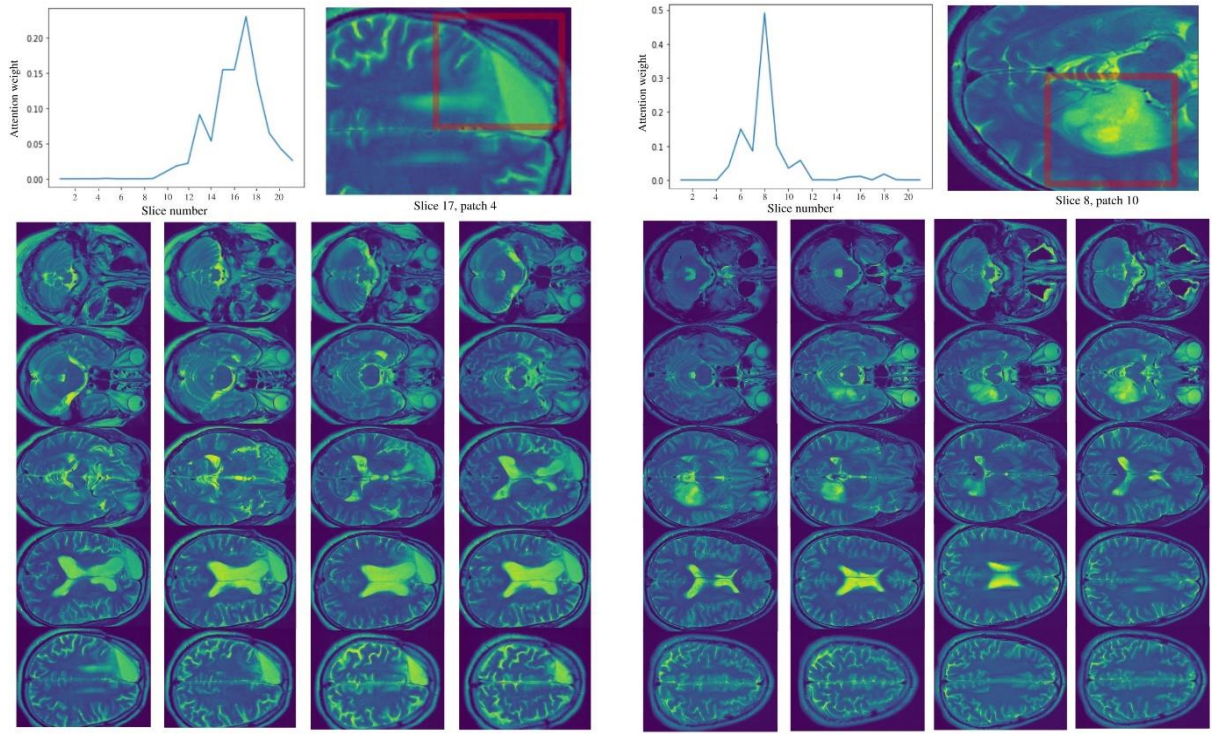

**Figure 7:** *Highest weighted patch within the highest weighted slice for two test set examples, demonstrating that the model can detect abnormalities throughout the brain.*

## 6. Discussion

The models presented in this work exploit the hierarchical nature of neuroimaging data to accurately classify a range of glioma-related abnormalities. By doing so it was shown that they outperform simpler models which treat separate sequences, slices, or intra-slice regions equally. Crucially, this hierarchical approach provides an intuitive form of model interpretability. By visualising patch-HAN's attention weights, it is possible to localise abnormalities both across and within individual slices, while sequence-HAN's attention weights provide inter-slice localisation while additionally providing importance scores for each imaging sequence. Critically, this is achieved without the need for slice- or pixel-level annotation during training, requiring only a series-level label which is applied to all slices. As such, our approach lends itself to training on large-scale retrospective hospital image collections, and is well suited for use as part of a semi-autonomous triage system. In the future, we wish to extend the model to allow incorporation of non-imaging data such as patient clinical history. This can be extracted from the free-text report that accompanies images on PACS and embedded into a machine-readable representation (e.g., using ALARM, BioBERT (Lee et al., 2019). ClinicalBERT (Alsentzer et al., 2019) language models) and introduced as an additional hierarchy. Furthermore, we plan to test the model on a range of abnormalities where sequences other than $T_2$-weighted images are important, for example, acute infarct cases for which DWI and apparent diffusion coefficient (ADC) maps appear particularly discriminatory. In the future, we also plan to combine the two models presented here to allow simultaneous sequence, slice, and region importance scores to be determined, However, for this work we were limited to a single 11 GB GPU which isn't sufficient for training this larger model.

## 7. Conclusion

In this work we introduced a hierarchical attention network to analyse real-world non-volumetric clinical MRI data, demonstrating that this hierarchical treatment of neuroimaging data leads to gains in model performance, while coarsely localising the abnormality both across and within individual image slices. As such, the model is suitable for use as part of a semi-automated triage system, where both model accuracy and interpretability are important.